\documentclass[pra,amsmath,showpacs,twocolumn,letterpaper]{revtex4}
\usepackage{times}
\usepackage{graphicx}

\usepackage{amsmath}
\usepackage{amssymb}
\usepackage{marvosym}
\usepackage{dsfont}
\usepackage{bezier,color,epsfig}

\begin{document}
\title{
How to STIRAP a vortex
} \author{G.~Nandi}
\email{gerrit.nandi@physik.uni-ulm.de}
\author{R.~Walser}
\author{W.~P.~Schleich}
\affiliation{Universit\"at Ulm, Abteilung f\"ur Quantenphysik, 
  D-89069 Ulm, Germany}
\date{\today, to be submitted to PRA}

\begin{abstract}
  We examine a scheme for the optical creation of a superfluid vortex in a
  trapped Bose-Einstein condensate (BEC), using the stimulated Raman adiabatic
  passage (STIRAP) technique. By exposing an oblate, axis-symmetric condensate
  to two co-propagating laser pulses, one can transfer external angular
  momentum from the light field to the matter wave, if one of the beams is the
  fundamental Gaussian mode and the other is a Gauss-Laguerre mode of angular
  momentum $1\hbar$.  We demonstrate the complete transfer efficiency by
  numerical integration of the multi-component Gross-Pitaevskii equation and
  explain the results with an intuitive and accurate approximation within the
  Thomas-Fermi limit. In addition, we discuss residual excitations (breathing
  modes) which occur in the two-dimensional regime and present the Bogoliubov
  excitation spectrum.
\end{abstract}

\maketitle

\section{Introduction}

Ultra-cold atomic gases have provided us with novel physical systems that
exhibit all genuine many-body phenomena known from traditional condensed
matter physics and, still, admit all the superior coherent control tools used
in quantum optics.  After the experimental realization of BEC itself,
tremendous efforts were focused on the creation of topological and solitary
excitations of condensates (for a current review see Ref.~\cite{fetter2002}).
Especially alternative methods for the creation of vortices have stirred the
minds, as the traditional ``rotating-the-squeezed-bucket'' procedure was not
successful, initially.  Thus, the first fruitful proposal to create a vortex
involved a rapidly rotating Gaussian laser beam entangling the external motion
with internal state Rabi-oscillations \cite{williams,mathews}.  Later,
condensates were stirred up mechanically
\cite{madison00a,madison01,ketterle01} and evaporative spin-up techniques
created vorticity \cite{haljan01} and recently giant vortices \cite{engels03}
could be created. Due to larger asymmetries in the trapping potentials that
can be achieved nowadays, vortices are now predominantly created with the
stirring method and fascinating Abrikosov lattices containing up to 300 vortices have
been made \cite{madison00b,ketterle01,cornell04}.  Alternatively, prospects
for creating vortices by optical phase imprinting \cite{lewenstein99} were
investigated (in analogy to the successful soliton experiment
\cite{denschlag00}) and applying magnetic interactions were considered
\cite{pu00}. Moreover, there have been ideas to create vorticity by sweeping a
laser beam on a spiraling trajectory across the trap, inducing a Landau-Zener
transition between the irrotational and the rotational state \cite{damski}.

On the other hand, the transfer of angular momentum from an optical field to a
macroscopic rigid body or an atomic particle has also a long standing
tradition in quantum mechanics. The first proof that circularly polarized
light carries angular momentum dates back to Beth's original experiment of
1936 \cite{beth}. More recently, due to the availability of Gauss-Laguerre
laser beams with well defined external angular momentum \cite{miloni,vanenk94}
it is possible to use them as optical tweezers and twisters
\cite{allen92,he95,kuppens98}.  Surprisingly, even the transfer of angular
momentum to ultra-sonic waves in fluids can be achieved that way
\cite{marte03}. In the context of a BEC, using the angular momentum of light
to create a doubly charged vortex has been proposed \cite{marzlin},
$\pi$-pulses in Raman type transitions were examined \cite{bolda98} and an
adiabatic passage to a vortex state was investigated by changing the
two-photon detuning of an effective two-level system \cite{dum98}.


In this paper, we will examine the transfer of external angular momentum of
light to the matter wave with the help of a stimulated Raman adiabatic passage
(STIRAP) \cite{bergmann}. The basic effect relies on a quantum mechanical
interference between two ground states and gives rise to a multitude of
physical phenomena, e.g., dark resonances in optical spectroscopy
\cite{arimondo76}, velocity selective coherent population trapping (VSCPT)
\cite{arimondo96}, a drastic modification of the optical index of refraction
(EIT) of normal \cite{harris90} and BE condensed systems
\cite{hau99,fleischhauer03}, as well as constructive procedures to prepare
\cite{parkins93} and readout quantum states of atomic beams and optical
cavities \cite{walser96}.

This article is organized as follows: In the Sec. II we will develop a scheme
for creating a vortex in a BEC using the STIRAP method. In analogy to
single-particle physics, it is possible to derive the relevant three-component
Gross-Pitaevskii equation. In Sec. III we will present the results of numerical
calculations that are in good agreement with a simple analytical approximation
within the Thomas-Fermi limit. In addition, we will discuss the physics of the
remaining residual excitations in terms of the "breathing modes" of a quasi
two-dimensional system. In particular, we will calculate the radial Bogoliubov
excitation spectrum of a condensate in the vortex state with angular momentum
$1 \hbar$.  Finally, we will summarize our results and conclude in Sec. IV.

\section{STIRAP in a Bose-Einstein condensed gas}

\begin{figure}[h]
  \begin{center} 
    \includegraphics[width=8.5cm]{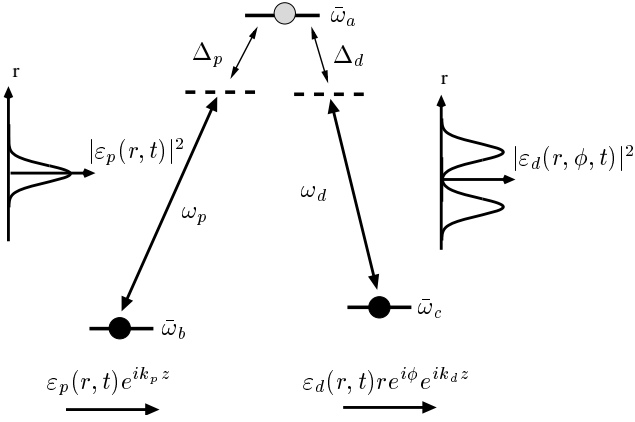}
    \caption{\label{lambda}A $\Lambda$-system exposed to two 
      co-propagating laser pulses in two-photon resonance.  The atomic
      transition frequencies are Doppler-shifted with respect to the rest
      frame i.e., $\bar{\omega}_a=\omega_a+\hbar k_p^2/2M$,
      $\bar{\omega}_b=\omega_b$, $\bar{\omega}_c=\omega_a+\hbar
      (k_p-k_d)^2/2M$.  Only one Gauss-Laguerre laser beam ($\epsilon_d$)
      carries $1 \hbar$ of angular momentum. The individual detunings from the
      excited state are $\Delta_p=\bar{\omega}_a-\bar{\omega}_b-\omega_p$,
      $\Delta_d=\bar{\omega}_a-\bar{\omega}_c-\omega_d$. }
\end{center}
\end{figure}

The STIRAP method is now applied to a trapped BEC of three-level atoms in a
$\Lambda$-type configuration shown in Fig.~\ref{lambda}. The two internal
electronic ground states, e.g. the hyperfine levels of an alkali atom
\cite{matthews98, hall98,hau99}, are denoted by $|b \rangle$ and $|c \rangle$, and by
absorbing an optical photon, one reaches the excited state $|a\rangle$,
respectively.  The condensate is confined spatially by an oblate,
axis-symmetric harmonic potential, which reads in cylindrical coordinates $(r,
\phi, z)$
\begin{equation}
\label{trap}
V(r,z)=\frac{1}{2}M( \omega^2 r^2+\omega_z^2 z^2),
\end{equation}
where $M$ denotes the single-particle mass. By choosing the radial trapping
frequency much less than the longitudinal frequency, i.\thinspace{}e.,
$\omega\ll \omega_z$, one can confine the motion effectively to the radial
component.

Now, we will expose the dilute atomic gas to two co-propagating traveling
"pump" and "dump" laser pulses,
\begin{eqnarray}
\label{beamp}
\mathbf{E}_{p}(z,t)&=&
\boldsymbol{\lambda}_{p}\,\varepsilon_{p}(t)\,
e^{-i(\omega_{p}t-k_{p}z)}+\text{c.c.},
\\
\label{beamd}
\mathbf{E}_{d}(r,\phi,z,t)&=&
\boldsymbol{\lambda} _{d}\,\varepsilon _{d}(r,\phi,t)\,
e^{-i(\omega _{d}t-k_{d}z)}+\text{c.c.},
\end{eqnarray}
where $\boldsymbol{\lambda}_p$ and $\boldsymbol{\lambda}_d$ denote the
corresponding polarization vectors.  The slowly varying laser beam envelopes
$\varepsilon_{p}$, $\varepsilon_{d}$ have a non-trivial temporal and spatial
structure
\begin{eqnarray}
\label{epsilonp}
\varepsilon_p(r,t)&\approx& \varepsilon_p(t) = \varepsilon_p \,
e^{-(t-\tau)^2/2\,d^2},
\\
\label{epsilond}
\varepsilon_d(r,\phi,t)&\approx&\varepsilon_d \, r \,e^{i \phi} 
\, e^{-t^2/2\,d^2}.
\end{eqnarray}
For the pump pulse, we choose the fundamental Gauss-Laguerre (GL) laser mode
\cite{miloni} with a spatial width larger than the BEC size and a temporal
Gaussian turn-on shape with the width $d$. Numerical simulations prove that in
this limit the spatial Gaussian envelope can be disregarded altogether in Eqs.
(\ref{epsilonp},\ref{epsilond}). This pulse reaches its maximum intensity at
some time $\tau>0$.  In order to transfer orbital angular momentum from the
light beam to the matter-wave, we pick the first excited GL mode that carries
external angular momentum, the so called "doughnut-mode", for the dump beam
\cite{vanenk94,allen92,marzlin,bolda98}.  While spatial extension and temporal
duration $d$ can be set equal in both pulses, it is crucial that the dump beam
reaches its maximum intensity at $t=0$, first.  This "counter-intuitive" pulse
sequence is the key of the STIRAP procedure and achieves an efficient
adiabatic passage for linear, dissipative quantum systems
\cite{arimondo96,bergmann,walser96}.  A full population transfer can be
reached if the field amplitudes satisfy the conditions
\begin{eqnarray}
\label{fp1}
\lim_{t \rightarrow -\infty}  
  \frac{\varepsilon_p(t)}{\varepsilon_d(t)} &=&0,
\\
\label{fp2}
\lim_{t \rightarrow +\infty}  
  \frac{\varepsilon_p(t)}{\varepsilon_d(t)} &=&\infty,
\end{eqnarray}
which is guaranteed by the pulse sequence given in Eqs.~(\ref{epsilonp},
\ref{epsilond}).

Far below the transition temperature $T\ll T_{BEC}$, one can describe a
multi-component BEC effectively within a simple mean-field picture
\cite{dalfovo,cirac1}. Thus, we introduce a three component state vector
$\boldsymbol{\Psi}(\mathbf{r},t)$, that represents the components of the
macroscopic atomic matter-wave. For the time evolution of this multi-level
state vector, one can derive a generalized Gross--Pitaevskii (GP) mean-field
equation \cite{walser01}
\begin{eqnarray}
\boldsymbol{\Psi}(\mathbf{r},t) &=& (\Psi _{a}(\mathbf{r},t),
\Psi _{b}(\mathbf{r},t), \Psi_{c}(\mathbf{r},t))^T,
\\
\label{GP}
i \hbar\, \partial _{t} \boldsymbol{\Psi}(\mathbf{r},t) &=& H(t) 
\,\boldsymbol{\Psi}(\mathbf{r},t).
\end{eqnarray}
$\boldsymbol{\Psi}(\mathbf{r},t)$ is normalized  to the total particle number
\begin{equation}
  N=\int \text{d}^3r \left( |\Psi_a(\mathbf{r},t)|^2 + 
    |\Psi_b(\mathbf{r},t)|^2 
    + |\Psi_c(\mathbf{r},t)|^2 \right),
\end{equation} 
in the BEC.  Due to the unitarity of Eq.~(\ref{GP}), this particle number is
conserved at all times. However, as we use explicitly time dependent laser
fields, the energy of the system can change (see Sec. III B).

The internal structure of the Hamiltonian is quite easy to understand, as it
follows straight from the single-particle physics that rules the dynamics of
the dilute gas interacting with light.  In Fig.~\ref{lambda}, we have depicted
the optical dipole transition scheme for a $\Lambda$-type atom.  Within the
standard rotating-wave approximation of quantum optics \cite{schleich}, one
finds for the internal state Hamiltonian
\begin{eqnarray}
\label{Hint}
H(t)/\hbar&=&\begin{pmatrix}
h_a+\Delta& \Omega_p(t)& \Omega_d(r,\phi,t)\\
\Omega_p^\ast(t)& h_b+\delta& 0\\
\Omega_d^\ast(r,\phi,t)&0&h_c-\delta
\end{pmatrix}.
\end{eqnarray}
The Rabi frequencies $\Omega _{p}(t)=\varepsilon _{p}(t) d_{ba}/\hbar$ and
analogously $\Omega _{d}(r,\phi,t)=\varepsilon_{d}(r,\phi,t)d_{ca}/\hbar$,
measure how well the photon field couples to the electronic transition and are
proportional to the atomic dipole moments $d_{ba}$, $d_{ca}$.  The remaining
parameters are the Raman detuning $\Delta=(\Delta_p+\Delta_d)/2$ and the
two-photon detuning $\delta=(\Delta_d-\Delta_p)/2$, which refer to the
individual detunings $\Delta_{p}$, $\Delta_{d}$, of laser the frequency and
the Doppler-shifted electronic transition frequency.  In order to achieve the
optimal STIRAP performance, we will assume a two-photon resonance condition
$\delta=0$ later, and pick a non-vanishing detuning $\Delta$ in order to avoid
detrimental spontaneous emissions, which would disrupt the coherent evolution.
In the preceeding derivation of Eq.~(\ref{Hint}), we have also tacitly adopted
co-moving and co-rotating reference frames such that
\begin{eqnarray}
\bar{\boldsymbol{\Psi}}(\mathbf{r},t)&=&
\left(
  e^{i[(\bar{\omega}_a-\Delta)t-k_p z]}\Psi_a(\mathbf{r},t),
  e^{i(\bar{\omega}_b-\delta)t} \Psi_b(\mathbf{r},t),\right.\nonumber\\
&&\left. e^{i[(\bar{\omega}_c+\delta)t-(k_p-k_d)z]} 
  \Psi_c(\mathbf{r},t)\right)^T,
\end{eqnarray}
in order to strip off the trivial plane wave character of the beams, see
Eqs.~(\ref{beamp},\ref{beamd}). To keep the notation simple, we will drop the
bar over the state vector in the following discussion.

For the external motion of the multi-component gas, we want to assume that all
components are confined by the same harmonic potential, cf. Eq.~(\ref{trap}).
This is not a stringent requirement for the excited state component $\Psi_a$,
as particles will only rarely occupy this level and move through it quickly.
Due to this low occupancy, it is also not necessary to consider any mean-field
shifts arising from self-interaction, or the motion through the remaining
components.  Consequently, we can simply use the bare trap Hamiltonian with a
Doppler shift
\begin{eqnarray}
\hbar h_a&=&\frac{\mathbf{p}^{2}}{2M}+V(r,z)+\hbar k_p \frac{p_z}{M}.
\end{eqnarray}
For the ground state components, we will only consider mean-field shifted
energy contributions that arise from elastic collisions and denote the inter-
and intra-species scattering lengths by $\{a_{bb},a_{bc},a_{cb},a_{cc}\}$. Hence,
these components read
\begin{gather}
  \hbar h_b=\frac{\mathbf{p}^{2}}{2M}+V(r,z)+ \frac{4\pi \hbar^2}{M}
  (a_{bb}\left| \Psi _{b}\right| ^{2}+
  a_{bc}\left| \Psi _{c}\right| ^{2}),\\
  \hbar h_c=\frac{\mathbf{p}^{2}}{2M}+V(r,z)+
  \hbar (k_p-k_d) \frac{p_z}{M} + \nonumber\\
  \frac{4\pi \hbar^2}{M} ( a_{cc}\left| \Psi _{c}\right| ^{2}+ a_{cb}\left|
    \Psi _{b}\right| ^{2}).
\end{gather}

The optical absorption-emission cycle imparts angular, as well as linear
momentum onto the final state matter-wave $\Psi_c$. However, linear momentum
cannot be conserved in a trapped system, and it would lead to a sloshing
motion along the z-direction.  This can be suppressed either, by choosing
equal laser frequencies and photon momenta, or by squeezing the trapping
potential into a very oblate configuration such that $\beta =
\omega_z/\omega\gg1$. This effectively "freezes" the longitudinal motion due
to an energy selection argument.

The later situation leads to a more stable configuration and is favorable.
Thus, we are able to approximate the state vector 
\begin{eqnarray}
  \boldsymbol{ \Psi}(\mathbf{r},t) &=&
  \left(
    \psi _{a}(r,t),
    \psi _{b}(r,t),
    e^{-i\phi }\,\psi _{c}(r,t)\right)^T \varphi _{0}(z,t),\nonumber\\
\end{eqnarray}
by factorizing it into a radial part, which is normalized to the number of particles in the BEC,
i.e.
\begin{equation}
\label{number_rad}
N = 2 \pi \int_0^{\infty} \text{d}r \, r \, \left( |\psi_a|^2 + |\psi_b|^2 +|\psi_c|^2 \right),
\end{equation}
and the ground state 
\begin{equation}
\varphi_0(z,t) = (\beta/\pi) ^{1/4}e^{-i\beta t/2}e^{-\beta z^{2}/2}
\end{equation}
of the one-dimensional harmonic oscillator in $z$--direction, which is
normalized to one, i.e.
\begin{equation} 
  \int_{-\infty}^{\infty} \text{d}z \, |\varphi_0|^2 = 1 . 
\end{equation}
Please note that the third component now carries the mechanical angular
momentum of $1\hbar$ per particle. After projecting the state vector along the
z-direction, we finally arrive at an effective Hamiltonian in the radial
direction.  Using the natural units of the transverse harmonic oscillator
(time unit $T_{\text{HO}}=2\pi/\omega$, length scale
$a_{\text{HO}}=\sqrt{\hbar/M \omega}$), one finds
\begin{gather}
\label{GPreduc}
i \partial _{t} \boldsymbol{\psi}(r,t) = H(t) \,\boldsymbol{\psi},\\
H(t)=\begin{pmatrix}
  h^{(0)} +\Delta& \Omega_p(t)& r \, \Omega_d(t)\\
  \Omega_p^\ast(t)&h^{(0)} + \delta + \kappa n(r)& 0\\
  r \,\Omega_d^\ast(t)&0&h^{(1)} - \delta + \kappa n(r)
\end{pmatrix},
\end{gather}
where $\kappa=\sqrt{8 \pi \beta} \, a$ and all scattering lengths are equal to
$a$. This assumption holds well for $^{87}\text{Rb}$ (see below). The particle
density is denoted by $n(r)=|\psi_b(r)|^2+ |\psi_c(r)|^2$, and $h^{(m)}$
represents a two-dimensional radial harmonic oscillator Hamiltonian
\begin{equation}
\label{hm}
h^{(m)}=-\frac{1}{2}(
\partial_r^2+\frac{1}{r}\partial_r-\frac{m^2}{r^2}-r^2)
\end{equation}
in an angular momentum manifold with $m\lessgtr 0$.
\section{Results and discussion}
In the following, we will be more specific and choose the typical scattering
parameters for $^{87}\text{Rb}$ \cite{matthews98, hall98}. To simplify
matters, we will also assume that the self-scattering and cross-component
scattering lengths are equal: $a_{bb}=a_{cc}=a_{bc}=a=110$ Bohr radii ($a_0$).
This turns out to be a very robust approximation and we comment on this later.
By implementing a numerical algorithm for solving the three-component
Gross-Pitaevskii equation Eq.~(\ref{GPreduc}), we can show that a quantized
vortex ($m=1$) is building up. For the trap frequencies we choose $\omega=2
\pi \cdot 10\,\text{s}^{-1}$ and $\omega_z=2 \pi \cdot 1000 \,\text{s}^{-1}$,
respectively.  With an atomic mass of $M=86.91$ amu, we find a harmonic
oscillator size $a_{\text{HO}}=\sqrt{\hbar/M \omega}=3.41 \,\mu$m. In this
spatial unit, we get for the scaled coupling constant $\kappa = \sqrt{8 \pi
  \beta} a=0.0855$.  We will consider an atomic ensemble with $N=10000$
particles. From a simple Thomas-Fermi approximation (see Eq.~(\ref{muTFlone}))
one finds estimates for the condensate radius $R_{\text{TF}}=\sqrt{2
  \mu_{\text{TF}}}=5.7$ with a chemical potential $\mu_{\text{TF}}=16.5$.  The
laser parameters are chosen as follows: $\tau=0.3$, $d=0.15$, $\Omega_p=200$,
$\Omega_d=200$, $\Delta=30$, and $\delta=0$ (two-photon resonance). In
practice, $\Delta$ will be of the order of at least $10^5$, but for the
simulations much lower values are sufficient.

In Fig.~2 the transfer efficiency is demonstrated. Initially the whole system
is prepared in the ground state $\left| b \right\rangle$. While the excited
state $\left| a \right\rangle$ is not populated significantly at any instant,
the population of state $\left| c \right\rangle$ is rising up. An almost 100\%
transfer to the vortex state is possible and the total number of particles in
the BEC is conserved as required by unitarity.  The time evolution of the
vortex state density is shown in Fig.~3. At some intermediate time ($t=0$)
high-frequency excitations appear. After the transfer is completed ($t=1$),
the shape of a vortex with angular momentum $1 \hbar$ is matched almost
perfectly with the stationary single charged vortex solution.  Still, some
minor excitations, i.\thinspace{}e., the "breathing mode" remains, which will
be discussed in Sec.~III~B.  In Fig.~4, we demonstrate the time evolution of
the density of the irrotational state $\left| b \right\rangle$.  Initially,
the atoms of the condensate are prepared in the ground state ($t=-1$).  This
population is decreasing during the STIRAP process and vanishes
afterwards.
\begin{figure}[h]
  \label{populations}
  \begin{center} 
    \vspace*{0pt}
    \includegraphics[width=8cm]{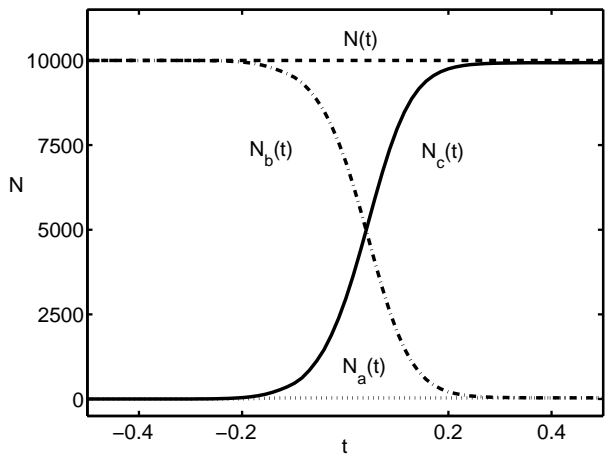}
    \caption{Time evolution of the individual level populations vs. time
      measured in natural units of $T_{\text{HO}}$. The dotted line depicts
      the population $N_a(t)$ of the excited state, the dashed-dotted and the
      solid lines represent $N_b(t)$ and $N_c(t)$, respectively, and the
      dashed line corresponds to the total number of particles in the system,
      which is conserved, i.e.  $N(t)=N_a(t)+N_b(t)+N_c(t)$.}
\end{center}
\end{figure}

While we have considered equal scattering lengths occurring accidentally in
$^{87}$Rb, this is not exactly valid and also not the generic situation
applicable to other condensed elements.  It could be imagined that the
imbalance of mean-field shifts deteriorates the transfer efficiency due to a
violation of the bare two-photon resonance condition. Still, one could apply a
time dependent detuning that rectifies this effect in order to achieve an
optimal adiabatic passage. Such a strategy has been proposed in
Refs.~\cite{marzlin,dum98}.

However, we have studied numerically the situation of unequal scattering
lengths without modifying the detunings and found no averse effects.  First,
we have examined the efficiency of the transfer procedure using the real
scattering data of the JILA experiment \cite{hall98, matthews98} for the
trapped hyperfine states $|b\rangle \equiv |F=1, m_F = -1\rangle$ and
$|c\rangle \equiv |2, 1\rangle$. Using the scattering parameters
$(a_{bb},a_{bc},a_{cc}) = 104 \,a_0\,(1.03,1,0.97)$, there is an almost
complete transfer to the rotational state possible, yielding essentially
identical curves as shown in Figs.~2-4.  However, there are slightly higher
excitations than in the idealized case.  Second, we find that even a drastic
variation of the scattering lengths does not spoil the transfer efficiency. It
only leads to a further increase of the ``breathing'' of the vortex. As an
example, we depict the results of the simulations for a vanishing
cross-component scattering length $(a_{bb},a_{bc},a_{cc}) = 104
\,a_0\,(1.03,0,0.97)$ in Figs.~5 and 8. Once more, these simulations confirm
the remarkable robustness of the STIRAP scheme.

\begin{figure}[h]
  \label{vortex1}
  \begin{center} 
    \vspace*{0pt}
    \includegraphics[width=8cm]{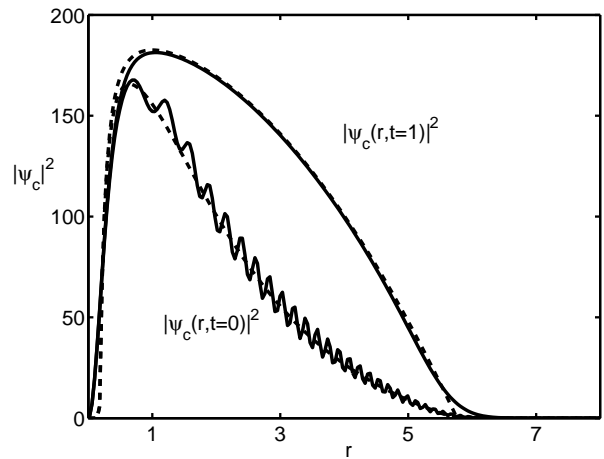}
    \caption{Time evolution of the vortex state density $|\psi_c(r,t)|^2$ vs. 
      radius $r$ measured in units of $a_{\text{HO}}$.  The solid lines depict
      the vortex state at an intermediate instant ($t=0$) and after the STIRAP
      process is completed ($t=1$).  In general, the instantaneous
      Thomas-Fermi approximation (dashed line) compares well with the exact
      results apart from minor excitations.}
\end{center}
\vspace{0.5cm}
\end{figure}

\subsection{The two-component Thomas--Fermi approximation}

The Thomas-Fermi (TF) limit is an extremely useful approximation for the
ground state of an interacting BEC.  Where applicable, it simply disregards
the kinetic energy contribution to the total system energy. This is also
possible for an excited vortex state. We will now consider this
TF-approximation for the two ground-state components, after adiabatically
eliminating the $\psi_a$ component, 
\begin{equation}
\psi_a(r,t)\approx-\frac{\Omega_p(t)}{\Delta} \psi_b-
\frac{r\, \Omega_d(t)}{\Delta}\psi_c,
\end{equation}
in order to obtain a simple estimate.
During the adiabatic passage, the irrotational state $\psi_b$ shall be
transformed coherently into the vortex state $\psi_c$. Thus, the relevant
Hamiltonian in the reduced two-state manifold reads
\begin{equation} 
\label{2compTF}
H_{TF}(t)=  \frac{r^2}{2}  + \kappa n(r)-\begin{pmatrix}
 \frac{\left|\Omega_p(t)\right|^2}{\Delta} 
  & \frac{\Omega_d(r,t) \Omega_p^\ast(t)}{\Delta}\\
  \frac{\Omega_d^\ast(r,t) \Omega_p(t)}{\Delta} & 
  \frac{\left|\Omega_d(r,t)\right|^2}{\Delta}-\frac{1}{2r^2}
\end{pmatrix}.
\end{equation}
Once more, we want to assume an adiabatic following condition to obtain a
stationary solution of $\boldsymbol{\psi}(r,t)=\exp{\left[-i \int_{-\infty}^t
  \text{d}\tau\mu_{TF} (\tau)\right]}\left(\psi_b(r),\psi_c(r)\right)^T$. By solving the simple
ensuing eigenvalue problem of Eq.~(\ref{2compTF}), we find for the chemical
potential in the vortex branch that
\begin{gather}
  \mu_{TF}= \frac{r^2}{2}+\kappa\,n_{TF}+  \delta\mu,\\
\delta\mu=\frac{1}{2\,\Delta} \left[ \frac{\Delta}{2\,r^2} -
  |\Omega_d|^2-|\Omega_p|^2 \right.\\
  \left. +\sqrt{
    |\Omega_d|^4 -
        2\,|\Omega_d|^2\,\left( \frac{\Delta}{2\,r^2} -
            |\Omega_p|^2 \right) + 
        {\left(
            \frac{\Delta}{2\,r^2} + |\Omega_p|^2 \right)
          }^2}\right].\nonumber
\end{gather}
To keep the notation simple, we have dropped all the spatial and temporal
arguments.  Vice versa, one has to determine the chemical potential $\mu_{TF}$
such that $N=2 \pi \int_{0}^{\infty} \text{d}r \, r \, n_{TF}$ is satisfied at
each instant of the adiabatic passage.  For the vortex component of the
corresponding eigenvector, one finds
\begin{gather}
\begin{pmatrix}
\psi_b \\
\psi_c 
\end{pmatrix}
=\sqrt{n_{TF}}
\begin{pmatrix}
 \cos{\theta} \\
 \sin{\theta}
\end{pmatrix}, \\
\tan{\theta}=\frac{\Re\left(\Omega_p^\ast \Omega_d\right)}{\Delta(
\frac{1}{2r^2}-|\Omega_d|^2-\delta\mu)}.
\end{gather}
It turns out that the chemical potential is almost constant during the
transfer process. Hence, we can choose the value of $\mu_{TF}$ for the
TF-solution of the vortex state, i.e.
\begin{equation}
\label{muTFlone}
\mu_{TF} = \mu_{TF}(\Omega_{p,d}=0) \approx \frac{r^2}{2} + \frac{1}{2 r^2} + \kappa n_{TF} = \text{const}.
\end{equation}

\begin{figure}[h]
  \label{novortex2}
  \begin{center} 
    \vspace*{0pt}
    \includegraphics[width=8cm]{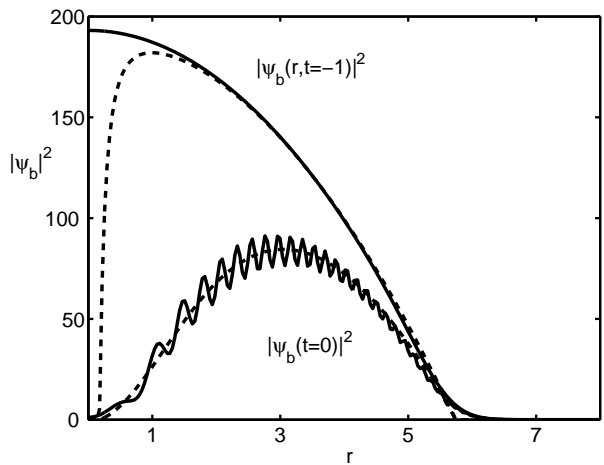}
    \caption{Time evolution of the irrotational state density 
      $|\psi_b(r)|^2$ vs. radius $r$ measured in units of $a_{\text{HO}}$.
      The solid lines depict the position density at the initial ($t=-1$) and
      at some intermediate time ($t=0$).  The instantaneous Thomas-Fermi
      approximation (dashed lines) matches the exact solution after the
      transfer of particles starts effectively ($t \geq 0$).}
\end{center}
\vspace{0.5cm}
\end{figure}

As shown in Fig.~3, the adiabatic evolution of these simple approximations
matches the exact numerical results for the vortex state well. It should
be pointed out that in our TF-approximation we did not neglect the centrifugal
term $1/2r^2$ in Eq.~(\ref{2compTF}), which is actually also part of the
kinetic energy. It is responsible for the vanishing density in the center of
the vortex core. This turns out to be the better approximation for $t \geq 0$,
while for $t<0$, dropping the centrifugal potential provides the more accurate
approximation because of the irrotational nature of the initial state.

\begin{figure}[h]
  \label{vortex1_a120}
  \begin{center} 
    \vspace*{0pt}
    \includegraphics[width=8cm]{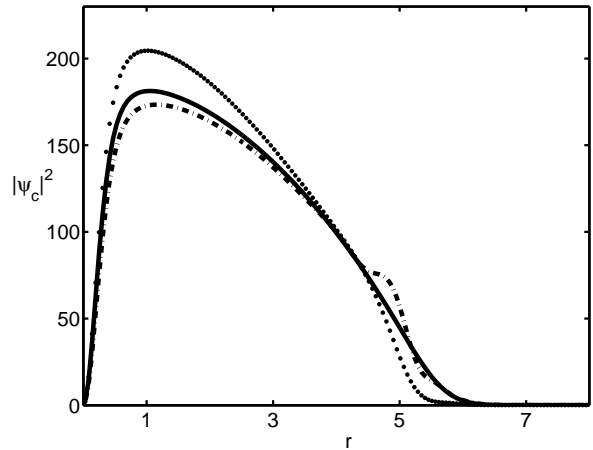}
    \caption{Time evolution of the vortex state density $|\psi_c(r,t)|^2$ vs. 
      radius $r$ measured in units of $a_{\text{HO}}$.  In this simulation the
      self-scattering lengths of $^{87}\text{Rb}$ were chosen, but $a_{bc}$
      was deliberately set to zero. This change gives rise to an increase of
      the residual excitations of the final vortex state while the transfer
      efficiency again is higher than 99$\%$.  The dashed and dotted lines
      show the ``breathing'' of the vortex state at different times $t$. As a
      reference, the solid line depicts the stationary vortex state.}
\end{center}
\vspace{0.5cm}
\end{figure}

\subsection{Total energy of the system}

The total energy of a BE condensed system can be obtained from the expectation
value of the microscopic Hamiltonian with respect to a symmetry-broken
ensemble \cite{blaizot,walser01}, thereby discarding higher order correlation
functions.  Unless time-translational symmetry is broken due to explicit time
dependencies of external fields, the total energy of the system must be
conserved.  Within these assumptions, the energy functional is given by
\begin{equation}
E = E_{\text{kin}} + E_{\text{TF}} + E_{\text{dip}},
\end{equation}
where the individual contributions are as follows
\begin{equation}
\label{Ekin}
E_{\text{kin}} =  \pi \int_0^{\infty} \text{d}r \ r \, \left[ |\partial_r \, \psi_a|^2 + |\partial_r \, \psi_b|^2+ |\partial_r \, \psi_c|^2\right],
\end{equation}
\begin{equation}
\begin{split}
\label{ETF}
E_{\text{TF}} =  \pi & \int_0^{\infty} \text{d}r \, r  \, \left[ 2\,\Delta\, |\psi_a|^2  
+ r^2\,(|\psi_a|^2+|\psi_b|^2 +|\psi_c|^2) \right.  \\
&\left. \ \phantom{+} + \frac{1}{r^2}\,|\psi_c|^2 + \kappa \,\left(|\psi_b|^2 + |\psi_c|^2 \right)^2\right]
\end{split}
\end{equation}
and
\begin{equation}
\label{Edip}
E_{\text{dip}} = 2 \pi \int_0^{\infty} \text{d}r \, r \, \left[ \psi_a^{\ast} \, \Omega_p \, \psi_b + \psi_a^{\ast} \, r \, \Omega_d \, \psi_c 
+ \, c.c. \right].
\end{equation}
Eq.~(\ref{Ekin}) denotes the radial kinetic energy $E_{\text{kin}}$, while
Eq.~(\ref{ETF}) represents all the energy arising from the detuning, the trap
potential, and the mean-field shifts, respectively. The dipole energy, which
is caused by the dipole coupling of the laser fields to the atoms, is given by
the expression in Eq.~(\ref{Edip}).

\begin{figure}[ht]
  \begin{center} 
    \includegraphics[width=8cm]{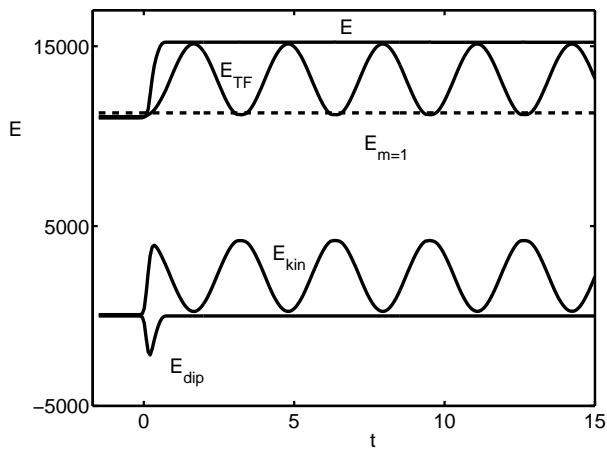}
    \caption{\label{energy}Total energy $E$ 
      (measured in units of $\hbar \omega$) as a function of the time $t$ (in
      natural units $T_{\text{HO}}$).  The different contributions to the
      total energy are shown in the plot.  $E_{m=1}$ denotes the energy of the
      condensate in the stationary vortex state.  Because of remaining
      excitations the potential and kinetic energy oscillate. However, the
      total energy is conserved before and after the transfer process.}
\end{center}
\end{figure}

The evolution of the total energy and its individual contributions is shown
during the non-equilibrium transfer in Fig.~\ref{energy}.  As expected, the
total energy is constant before and after the STIRAP process.  However, due to
an excitation of the breathing mode, the kinetic as well as the potential
energy exhibit complementary oscillations.  The dipole energy is negative, which
is well-known from the interaction of a two-level atom with a laser field.
This is due to the fact that the polarization of the atom is counteracting the
external electromagnetic field.

\subsection{Linear response of the system and breathing modes}


The numerical simulations show that a transfer of almost $100\%$ of the
particles into a vortex state is possible. Still, residual radial excitations
of the vortex state remain.  As the frequency of these radial excitation is
exactly twice the radial harmonic trapping frequency, $\varepsilon = 2 \, \omega$, it
must be related to the scaling symmetry of the two-dimensional system.  These
"breathing modes" have been studied first theoretically in
Refs.~\cite{pitaevskii1996,kagan} and experimentally in Refs.~\cite{chevy}.

The mechanism for exciting these modes arises from squeezing the harmonic
potential in time. In the present context of the STIRAP configuration, the
origin of such an additional potential can be understood from considering
Eq.~(\ref{2compTF}). During the turn-on of the laser fields, we induce an
ac-Stark-shift potential $|\Omega_d(r,t)|^2/\Delta$, which is proportional to
$r^2$.  While this effect is interesting by itself, it could be eliminated  
easily by an appropriate control of the trapping potential,
thus reducing the amount of excitations.

\begin{figure}[h]
  \begin{center} 
    \includegraphics[width=8cm]{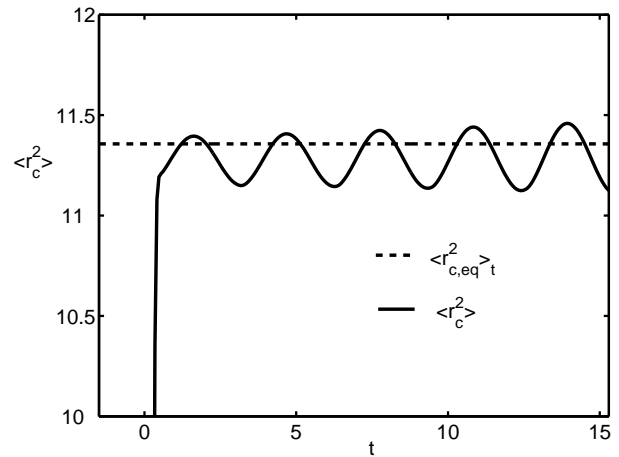}
    \caption{\label{r2}
      Mean squared radius $\langle r_c^2 \rangle_t$ of the vortex state vs.
      time (measured in units of $a_{\text{HO}}$ and $T_{\text{HO}}$,
      respectively).  Before the transfer process this state is not occupied.
      Afterwards the mean squared radius oscillates around the equilibrium
      value $\langle r^2_{c,eq.} \rangle$ (dashed line). The frequency of the
      excitation is exactly $2 \, \omega$.}
\end{center}
\end{figure}

\begin{figure}[h]
  \begin{center} 
    \includegraphics[width=8cm]{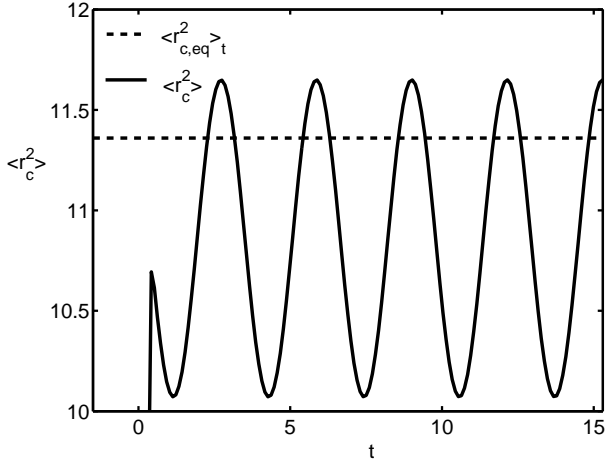}
    \caption{\label{r2_a120}
      Mean squared radius $\langle r_c^2 \rangle_t$ of the vortex state vs.
      time (measured in units of $a_{\text{HO}}$ and $T_{\text{HO}}$,
      respectively).  In this simulation the self-scattering lengths of
      $^{87}\text{Rb}$ were chosen, but $a_{bc}$ was deliberately set to zero.
      This change corresponds to an increase of the ``breathing'' of the final
      vortex state.}
\end{center}
\end{figure}
 
Moreover, a completely adiabatic transfer is limited by to two factors: On the
one hand a standard STIRAP process in a homogeneous gas requires an
adiabaticity condition $\Omega \tau \gg 1$, i.e. the time delay $\tau$ must be
sufficiently large. On the other hand $\tau$ cannot be chose arbitrarily large
because here we deal with a STIRAP process in a trap. Therefore, coupling to
external motion has to be taken into consideration for large time delays
$\tau$.

The residual radial excitations can be visualized by evaluating the 
mean squared radius
\begin{equation}
  \langle r_c^2 \rangle_t = \frac{2\pi}{N} \int_0^{\infty} \, \text{d}r \, r \, \psi_c(r,t)^* r^2 \psi_c(r,t),
\end{equation}
which is proportional to the potential energy of the vortex-component.  As
shown in Fig.~\ref{r2}, this quantity is oscillating with the frequency
$\varepsilon=2 \, \omega$, which corresponds to the breathing mode.

In addition, small radial excitations (i.e. the linear response) of a BEC can be
understood from Bogoliubov theory \cite{bogol47,fetter72}. Therefore,
we now consider the radial one-component GP-equation for the vortex-state with
angular momentum $1 \hbar$ that reads
\begin{equation}
i \partial_t \psi(r,t) = \left( h^{(1)} + \kappa \, |\psi(r,t)|^2 \right) \psi(r,t),
\end{equation}
with $h^{(1)}$ defined in Eq.~(\ref{hm}).
With the ansatz
\begin{equation}
\psi(r,t) = \text{e}^{-i \mu t} \left( \psi_{0}(r) + u(r) \, \text{e}^{-i \varepsilon t} + v^*(r) \, \text{e}^{i \varepsilon t} \right),
\end{equation}
where $\psi_{0}(r)$ is the wave function of the stationary vortex, the time evolution of which is
determined by the chemical potential $\mu$, and $u(r)$ and $v(r)$ denote the excitation modes with the normalization ($\varepsilon>0$)
\begin{equation}
2 \pi \int_{0}^{\infty} \text{d}r \ r \, \left( |u(r)|^2 - |v(r)|^2 \right) = 1.
\end{equation}
The spectrum can be calculated from the linear response eigenvalue problem
\begin{equation}
\label{bogolewert}
\begin{pmatrix}
h & \kappa \, \psi_0^2\\
-\kappa \, \psi_0^{\ast \, 2} & - h^{\ast} 
\end{pmatrix}
\begin{pmatrix}
u(r) \\
v(r) 
\end{pmatrix}= \varepsilon \begin{pmatrix}
u(r) \\
v(r)
\end{pmatrix} ,
\end{equation}
where
\begin{equation}
h = h^{(1)} - \mu + 2 \kappa |\psi_0(r)|^2.
\end{equation}

\begin{figure}[h]
  \begin{center} 
    \includegraphics[width=8cm]{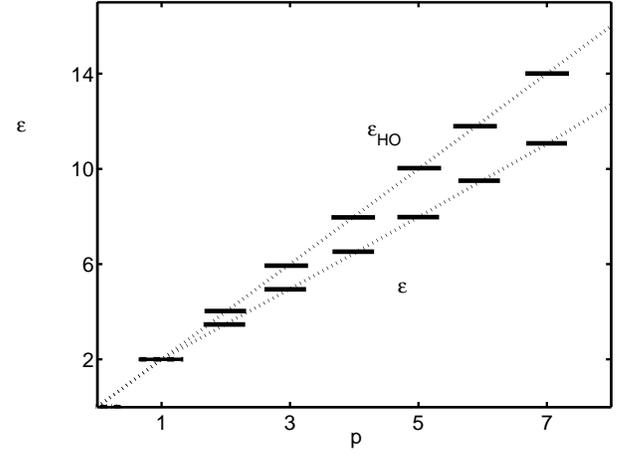}
    \caption{\label{enbog}
      Excitation frequencies $\varepsilon$ relative to the ground state
      frequency of a quasi-two-dimensional BEC in the vortex state ($|m|=1$)
      measured in units of $\omega$, in comparison to the
      two-dimensional harmonic oscillator with fixed $|m|=1$. For $p=0$ and
      $p=1$ the linear and nonlinear energies are identical. For $p \geq 2$
      they split, and the excitation energies of the BEC lie below those of
      the non-interacting gas.}
\end{center}
\end{figure}
\begin{figure}[h]
\begin{center} 
    \includegraphics[width=8cm]{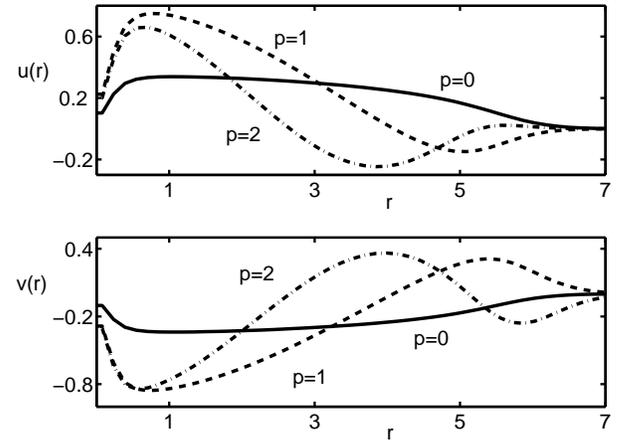}
    \caption{\label{moden}Excitation modes obtained by Bogoliubov theory. 
    The first three modes of $u(r)$ and $v(r)$ are plotted. The lowest mode ($p=0$) corresponds to the condensate wave-function itself
    (Goldstone-mode).}
\end{center}
\end{figure}

The results of our numerical calculations are shown in Fig.~\ref{enbog} and Fig.~\ref{moden}, being in agreement with earlier work
\cite{isoshima}. The lowest mode ($p=0$) corresponds to the condensate wave-function itself
(Goldstone-mode).
The frequency spectrum can be compared to spectrum of the well-known two-dimensional quantum-mechanical
harmonic oscillator, which is given by
\begin{equation}
\varepsilon_{HO} = 2p + |m| + 1,
\end{equation}
where $p \in \mathds{N}$ denotes the principal quantum number, $m \in \mathds{Z}$ denotes the angular momentum,
and, in our case, $|m|=1$ is fixed.

\section{conclusions}
In this paper, we have developed a novel scheme for the optical creation of
vortices in a trapped Bose-Einstein condensate, using the technique of
stimulated Raman adiabatic passage (STIRAP). In our model, we considered a BEC
of three-level atoms in a $\Lambda$-configuration of the electronic states,
which are coupled by two co-propagating laser pulses.  The aim was the
transfer of angular momentum, carried by one of the beams (Gauss-Laguerre
mode), to the BEC.  The underlying mechanism of STIRAP is analogous to
single-particle physics. In contrast to the latter case, we derived a
multi-component nonlinear Schroedinger equation (Gross-Pitaevskii equation),
using the mean-field approximation.  We presented results of numerical
simulations that apply to a BEC of $^{87}\text{Rb}$-atoms. For a suitable set
of laser parameters an almost $100\%$ transfer to the vortex state can be
achieved. These results can be understood with an intuitive and accurate
approximation within the Thomas-Fermi limit.  The occurrence of residual radial
excitations in the vortex state can be explained by so-called breathing
modes, which are specific for the two-dimensional regime and can be eliminated
by an appropriate control of the trap frequency. To confirm these explanation,
we have calculated the Bogoliubov excitation spectrum numerically.

\section*{Acknowledgments}
We would like to acknowledge fruitful discussions with Bruce W. Shore
and Karl-Peter Marzlin.


  
\bibliographystyle{prsty}

\end{document}